\documentclass[submission,copyright,creativecommons]{eptcs}
\providecommand{\event}{EXPRESS-SOS 2023} 

\usepackage[centerlast]{subfigure}
\usepackage{graphicx}

\usepackage{iftex}
\usepackage{listings}
\usepackage{color}
\usepackage{multicol}

\lstdefinelanguage{rebeca}{
  morekeywords={reactiveclass, knownrebecs, statevars, main, msgsrv, main, define, property, LTL, TCTL, boolean, int, shortint, byte, if, else, while, for, wait, msg, reset, set, self, false, true, now, after, delay, deadline, initial},
  otherkeywords={=>,<-,<\%,<:,>:,\#,@},
  sensitive=true,
  morecomment=[l]{//},
  morecomment=[n]{/*}{*/},
  morestring=[b]",
  morestring=[b]',
  morestring=[b]"""
}

\ifpdf
  \usepackage{underscore}         
  \usepackage[T1]{fontenc}        
\else
  \usepackage{breakurl}           
\fi

\title{Timed Actors and Their Formal Verification}
\author{Marjan Sirjani
\institute{M\"alardalen University\\ V\"asterås, Sweden}
\email{marjan.sirjani@mdu.se}
\and
Ehsan Khamespanah
\institute{University of Tehran\\
Tehran, Iran}
\email{e.khamespanah@ut.ac.ir}
}
\def\titlerunning{Timed Actors and Their Formal Verification}
\def\authorrunning{M. Sirjani, E. Khamespanah}
\begin{document}
\maketitle

\begin{abstract}
In this paper we review the actor-based language, Timed Rebeca, with a focus on its  formal semantics and formal verification techniques. 
Timed Rebeca can be used to model systems consisting of encapsulated
components which communicate by asynchronous message passing. Messages are put in the message buffer of the receiver actor and can be seen as events. Components react to these messages/events and execute the corresponding message/event handler.
Real-time features, like computation delay, network delay and periodic behavior, can be modeled in the language. We explain how both Floating-Time Transition System (FTTS) and common Timed Transition System (TTS) can be used as the semantics of such models and the basis for model checking.  We  use FTTS when we are interested in event-based properties, and it helps in state space reduction. For checking the properties based on the value of variables at certain point in time, we use the TTS semantics.
The model checking toolset supports schedulability analysis,  deadlock and queue-overflow check, and assertion based verification of Timed Rebeca models. TCTL  model checking based on TTS is also possible but is not integrated in the tool.

\end{abstract}

\section{Introduction}
Actors are introduced for modeling and implementation of distributed systems \cite{hewitt1977viewing,agha1986actors}. Timed Actors allow us to introduce timing constraints, and progress of time, and are most useful for modeling 
time-sensitive systems.
Timed Rebeca is one of the first timed actor languages with model checking support \cite{khamespanah2015timed}. 
Timed Rebeca restricts the modeller to a pure asynchronous actor-based
paradigm, where the structure of the model can represent the service oriented architecture,
while the computational model matches the network infrastructure \cite{DBLP:journals/corr/abs-1108-0228}.
In a different context, it may represent components of cyber-physical systems, where components are triggered by events put in their input buffers, or by time events \cite{Sirjanimath8071068}.
Timed Rebeca is equipped
with analysis techniques based on the standard semantics of timed systems, and also an innovative event-based semantics that is tailored for
timed actor models \cite{sirjani2016time}.

Timed Rebeca is an extension of the Reactive Object Language, Rebeca \cite{sirjani2004modeling}. 
It is reviewed and compared to a few other actor languages in a survey published in ACM Computing Surveys in 2017 \cite{boer2017survey}.
The very first ideas of Rebeca and its compositional verification is presented at AVoCS workshop in 2001 \cite{Avocs2001}.
Timed Rebeca, different formal semantics of it, and the model checking support are presented in multiple papers. Here we present an overall view and insight into different semantics and use a simple example to show the differences visually.

\section{Timed Rebeca}


A Timed Rebeca model mainly consists of a number of \textit{reactive class} definitions. These reactive classes define the behavior of the classes of the actors in the model. The model also has a \texttt{main} block that defines the instances of the actor classes. 

We use a simple Timed Rebeca model as an example to  explain the language features. In this example we consider two different actors. The first actor is able to handle three different tasks, named as \textit{job1}, \textit{job2}, and \textit{job3}. The second actor can only handle one task, named as \textit{job4}. 
%
The Timed Rebeca model of this example is shown in Listing~\ref{listing::rebeca-model}. there are two classes of actors: \texttt{Actor1} (lines 1-15) and \texttt{Actor2} (lines 17-27). The \texttt{main} block in lines 29-32 defines one instance of each class. 
Each reactive class has a number of \textit{state variables}, representing the local state of the actors. They may contain variables of basic data types, including booleans, integers, arrays, or references to other actors. To make the example model simple, none of the reactive classes of Listing~\ref{listing::rebeca-model} has any state variables. Each class can have a 
\textit{constructor}, which is used to initialize the created instances of the class by initializing the state variables, and start up running of the model by sending  
messages to itself or other actors.

In the Timed Rebeca model of Listing~\ref{listing::rebeca-model}, in the constructor of \texttt{Actor1} (line 3), the actor
sends itself a \texttt{job1} message. Each reactive class accepts a number of message types which are handled using \textit{message servers} (\texttt{msgsrv}). \texttt{Actor1} has three message servers, \texttt{job1} (lines 5-8), \texttt{job2} (lines 9-11), and \texttt{job3} (lines 12-14). Serving a message of type \texttt{job1} results in sending \texttt{job2} message to self which is put in the message buffer of itself only after passing 1 unit of time (modeled by using the \texttt{after} construct).  The  \texttt{deadline} construct denotes the deadline of the message to be handled, if at the time of handling the event the deadline is passed the model checking tool notifies that.
Then, there is a \texttt{delay} statement which models progress of time for 5 units of time, this can be used to model a computation delay. In the definition of the message servers, well-known program control structures can be used, including \textit{if-else} conditional statements, \textit{for} and \textit{while} loops, the definition of local variables, and assignments using usual arithmetic, logic, and comparative operators.

\begin{lstlisting}[language=rebeca, multicols=2, caption=A simple Timed Rebeca model with two actors., label=listing::rebeca-model]
reactiveclass Actor1(3) {
	Actor1() {
		self.job1();
	}
	msgsrv job1() {
		self.job2() after(1) deadline(10);
		delay(5);
	}
	msgsrv job2() {
 
	}
	msgsrv job3() {
		self.job3() after(1);
	}
}

reactiveclass Actor2(3) {
	knownrebecs {
		Actor1 a1;
	}
	Actor2() {
		self.job4() after(2);
	}
	msgsrv job4() {
		a1.job3() after(2) deadline(5);
	}
}

main {
	Actor1 actor1():();
	Actor2 actor2(actor1):();
}
\end{lstlisting}

In Timed Rebeca models, we assume that actors have local clocks which are synchronized throughout the model. Each message is tagged with a time stamp (called a time tag). 
We use a \texttt{delay(t)} statement to model the computation delay, and we use  \texttt{after(t)} in combination with a \texttt{send} message statement to model a network delay, or model a periodic event. When we use \texttt{after(t)} in  a \texttt{send} message statement it means that the time tag of the message when it is put in the queue of the receiver is the value of the local clock of the sender plus the value of \texttt{t}. The progress of time is forced by the \texttt{delay} statement. We assume that the local clock of  each actor is zero when the model starts execution, the local clock is increased by value of \texttt{t} if there is a \texttt{delay(t)} statement. A \texttt{send} statement with an  \texttt{after} does not cause an increase in the local time necessarily. The local time of the receiver actor is set to the time tag of the message when the actor picks the message, unless it is already greater than that. The latter situation means that the message  sits in the queue while the actor is busy executing another message, in this case the \texttt{after} construct does not cause progress of time.
The progress of time happens in the case that the time tag of the message is greater than the local time of the receiver actor, in this case the local time will be pushed forward.
In Timed Rebeca, messages are executed atomically and are not preempted.

\section{Different Semantics of Timed Rebeca}

\begin{figure*}
\centering
\subfigure[TTS of the 
Timed Rebeca model of Listing~\ref{listing::rebeca-model}]{
\label{fig::TTS-DTG-three-states}
  \centering
  \small{
   \includegraphics[width=.60\textwidth]{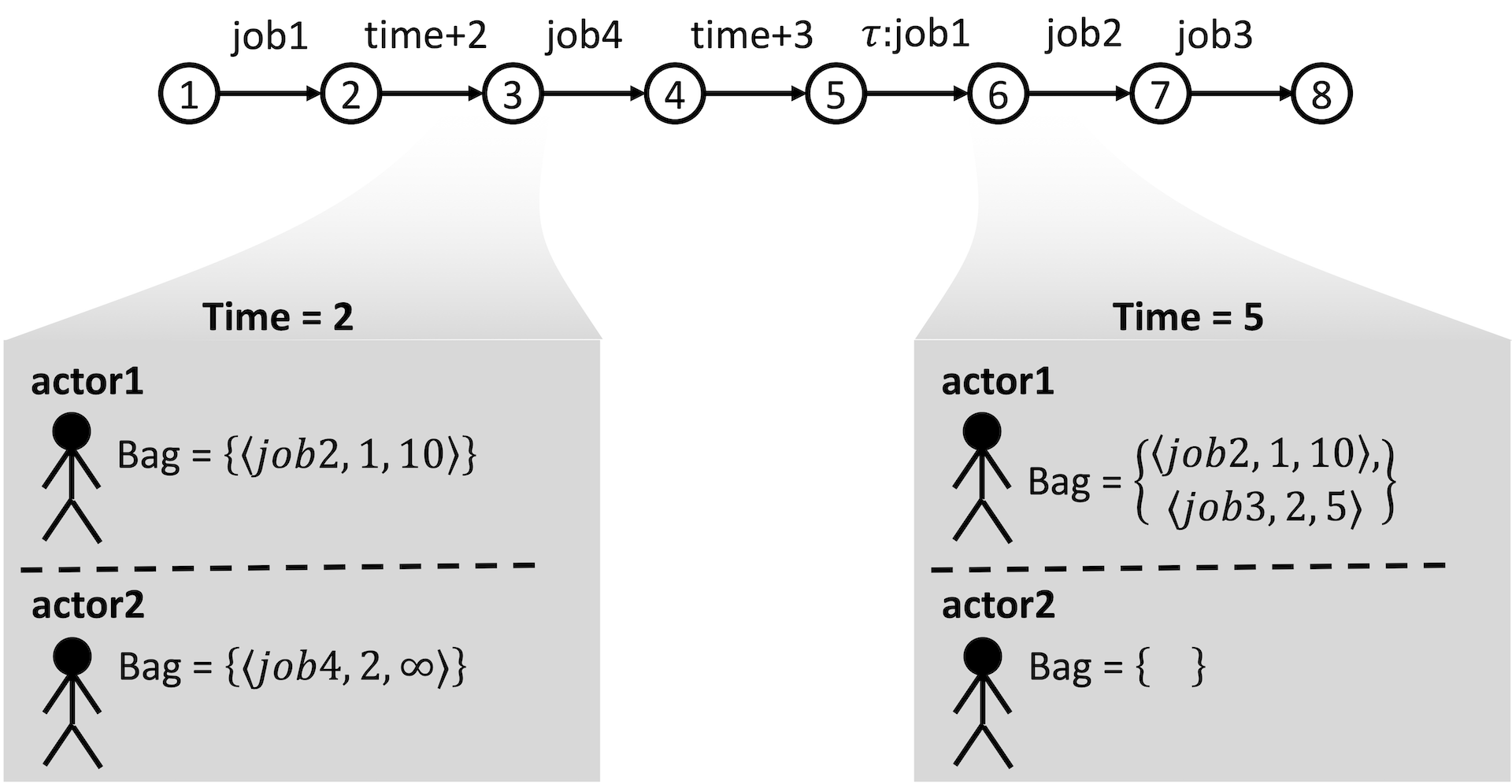}
  }
}
\qquad
\subfigure[FTTS of the Timed Rebeca model of Listing~\ref{listing::rebeca-model}. Each actor has its own local clock represented by Time. The value of local clock is considered in choosing the next transition.]{
\label{fig::jump-semantic}
  \centering
  \small{
   \includegraphics[width=.45\textwidth]{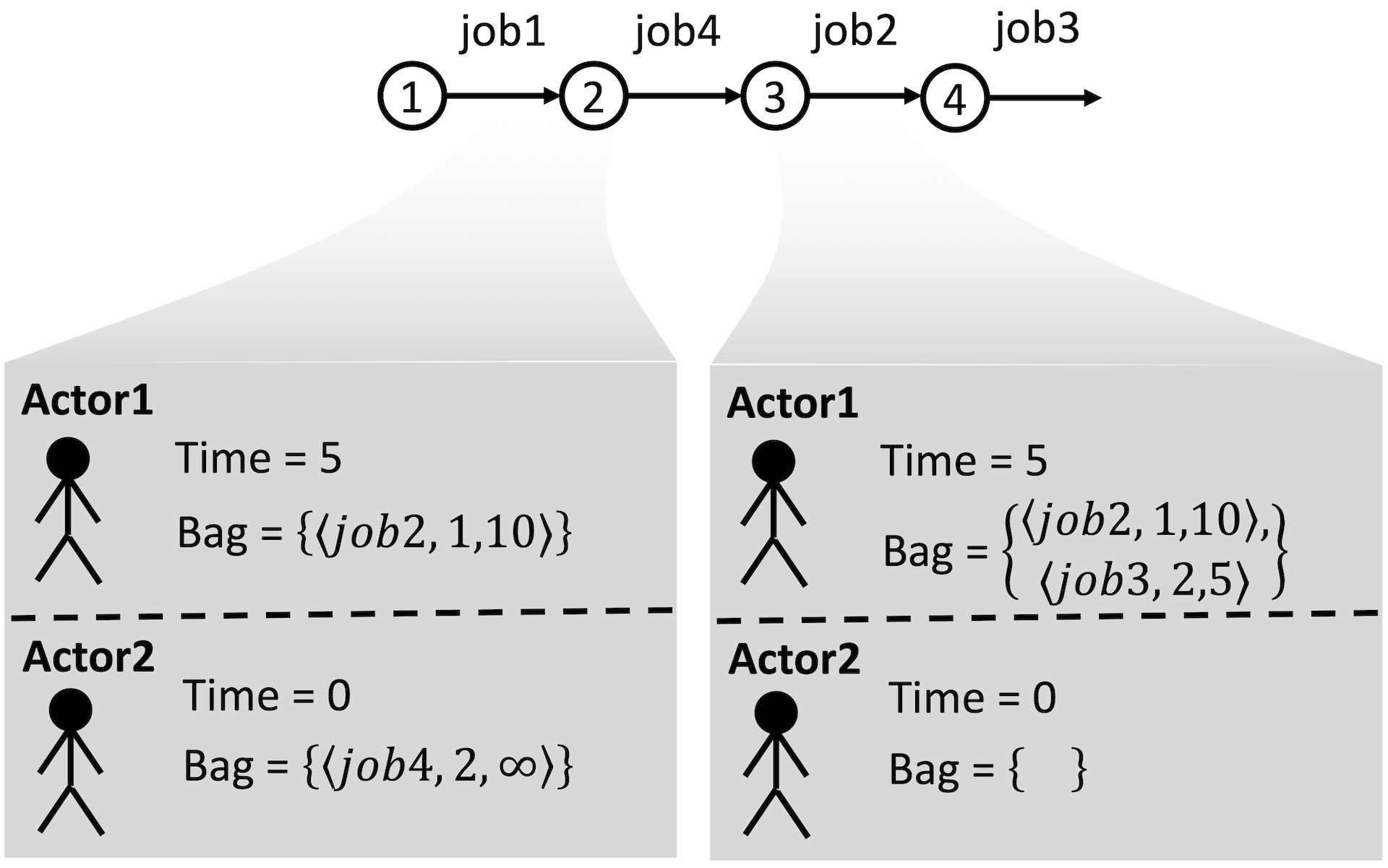}
  }
}
\qquad
\subfigure[Relaxed-FTTS of the Timed Rebeca model of Listing~\ref{listing::rebeca-model}. Each actor has its own local clock represented by Time. The message with the lowest time tag is chosen and the execution of its message server is the label of the next transition.]{
\label{fig::cont-semantic}
  \centering
  \small{
   \includegraphics[width=.45\textwidth]{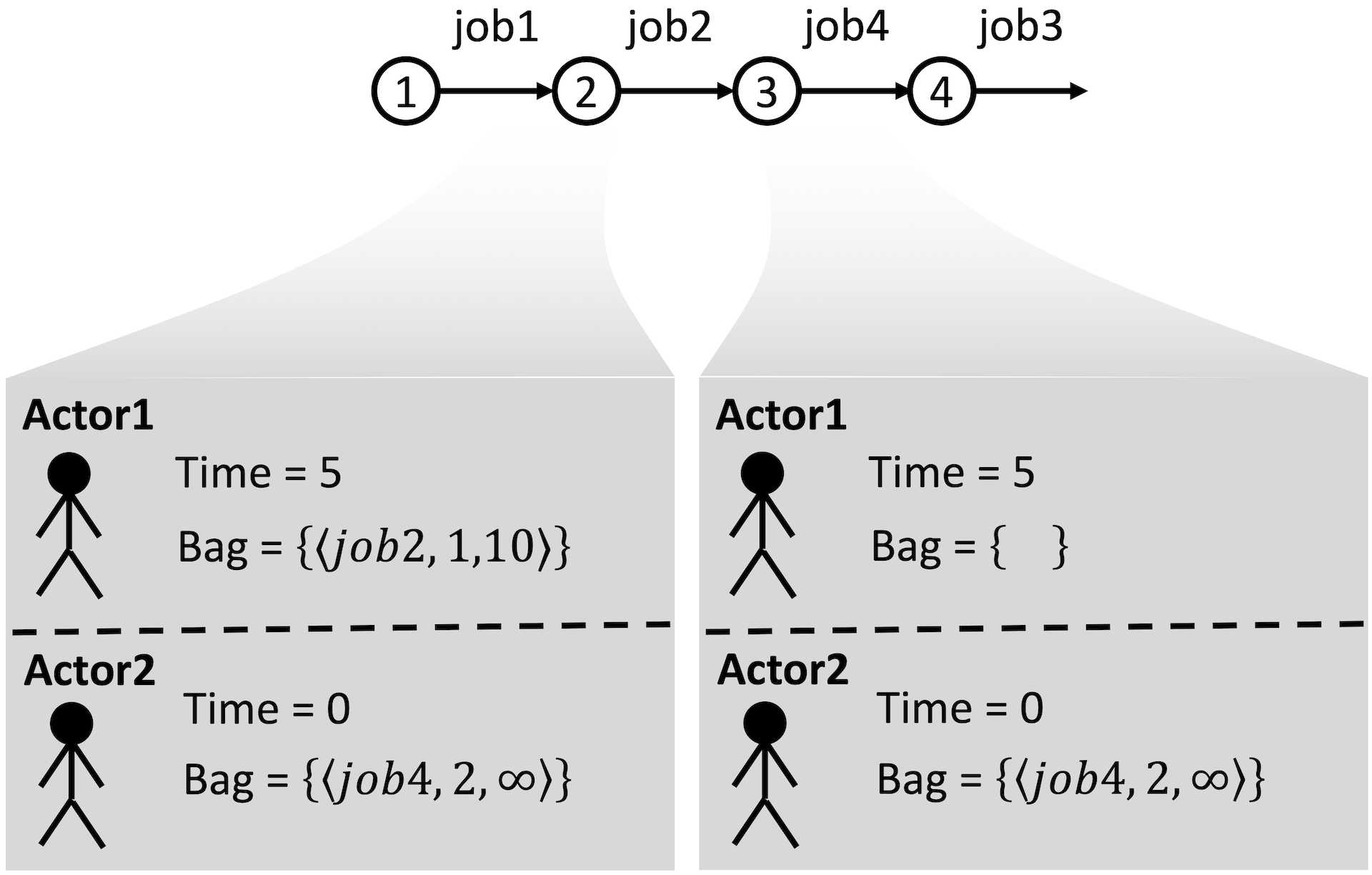}
  }
}
\caption{Comparing TTS, FTTS, and the relaxed form of FTTS for the Timed Rebeca model of Listing~\ref{listing::rebeca-model}.}
\label{fig::TTS-FTTS-RelFTTS}
\end{figure*}

We first introduced an event-based semantics for Timed Rebeca and used McErlang for simulation of Timed Rebeca models in \cite{DBLP:journals/corr/abs-1108-0228,DBLP:journals/scp/ReynissonSACJIS14}. In this semantics we focused on the object-based features of actors, encapsulation and information hiding, and decided on a coarse-grained semantics where serving a message (or handing a request or signal) are the only observable behavior of actors. 
We considered taking a message from top of the message queue and executing it as an observable action, and we called it an event. 
Note that by a message queue in Timed Rebeca, we mean a bag of messages where each message has a time tag of when the message is put in the buffer. Here, by ``top of the message queue of an actor'', we mean the message with the least time tag in the bag of messages targeted to that actor.
In defining the formal semantics of Timed Rebeca as a labeled transition system, we only have one type of label on the transitions, \textit{events}, which are taking messages and executing them.
In \cite{DBLP:conf/facs2/KhamespanahSVK15}, and its extended version \cite{khamespanah2015timed}, we introduced this event-based semantics of Timed Rebeca as Floating Time Transition System (FTTS) and compared it with the time semantics that is generally used for timed models (for example for Timed CCS \cite{10.5555/1324845}) where the transitions can be of the type of an event, progress of time, and a silent action.

Although we consider FTTS as the original and most fit semantics for Timed Actors, it may also be seen as a reduction technique in model checking. FTTS can give a significant reduction in state space compared to the standard Timed Transition System.
%
In \cite{khamespanah2015timed} we proved that there is an action-based weak bisimulation relation between FTTS and TTS of Timed Rebeca. 
Note that the focus here is on the labels on the transitions not on the values of variables in the states.

The semantics presented in \cite{DBLP:journals/scp/ReynissonSACJIS14}, is a relaxed form of FTTS in \cite{khamespanah2015timed} where in choosing the next step in a state we have a simpler policy.
The SOS rules of FTTS and the relaxed version are presented in \cite{khamespanah2015timed}  and \cite{DBLP:journals/scp/ReynissonSACJIS14}, respectively.
In each state, the SOS rule for the scheduler  chooses the next message in the bags of actors to be executed. In the relaxed form of FTTS, the scheduler simply chooses the message with the least time tag (targeted to any actor). In FTTS, the schedular considers the local clock of each actor as well.  For each actor, the maximum between the local clock  and the lowest time tag of the messages in the message bag of the actor is computed. Then among all the actors, the scheduler chooses the actor with the least of these amounts. The message on the top of the queue of this chosen actor will  be executed next.
Comparing to the standard TTS, the relaxed form of FTTS preserves the order of execution of messages of all actors if we consider the time tags of messages for ordering. The intuitive reason is that in Timed Rebeca we  consider a FIFO  policy for scheduling the messages in the message buffer, when we choose the message with the lowest time tag to be executed, it is guaranteed that from that point on, there will be no messages with a smaller time tag added to the message buffer (of any actor). So, the FIFO policy for serving messages can be correctly respected. The subtle point here is that the actor \textit{a} with the lowest time tag message \textit{m} may be busy when  message \textit{m} is sitting in its message buffer, in the meanwhile other messages from other actors may get the chance to be executed and send messages to  actor \textit{a}. Of course, the time tag of those messages will be greater than the time tag of message \textit{m}, but still we are losing the ``correct'' content of the message buffer of \textit{a} at some snapshots in time. By this observation, we moved to the FTTS semantics in \cite{khamespanah2015timed} where at any point in time, we have the correct content of the message buffer. Using this semantics we may choose to use other scheduling policies for messages (events) in the buffer, for example the \textit{earliest deadline first} policy.

In Figure \ref{fig::TTS-FTTS-RelFTTS}, we show parts of the the state transition system for Rebeca model in Listing \ref{listing::rebeca-model}. In this figure, we see how in TTS we may have three types of labels on the transitions, an event, time progress and $\tau$ (silent) transitions. In FTTS, in state 2 in Figure 1.b, the scheduler chooses the message \texttt{<job4,2,$\infty$>} while in the relaxed form of FTTS, in Figure 1.c, the scheduler chooses the message \texttt{<job2,1,10>}. The reason is that although the message with the lowest time tag is \texttt{<job2,1,10>}, with the time tag 1, the maximum between 1 and the value for the local clock of \texttt{Actor1} is 5.
The maximum between 2 
 (the time tag for message \texttt{<job4,2,$\infty$>} ) and the value for the local clock of \texttt{Actor2} (which is zero in this state) is 2.

\section{Model Checking Timed Rebeca Models}

\begin{figure*}
\centering
\subfigure[The state space for TTS-based semantics
]{
\label{fig::DTG-three-states}
  \centering
  \small{
   \includegraphics[width=.17\textwidth]{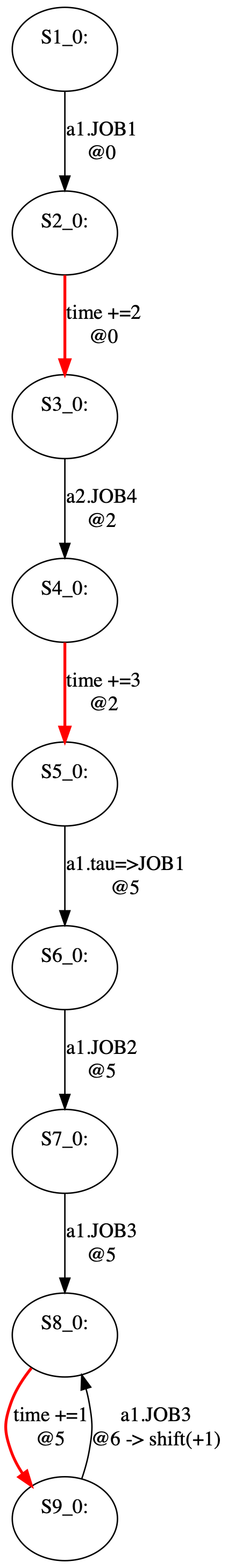}
  }
}
\qquad
\subfigure[The state space for FTTS-based semantics
]{
\label{fig::State-jump-semantic}
  \centering
  \small{
   \includegraphics[width=.2\textwidth]{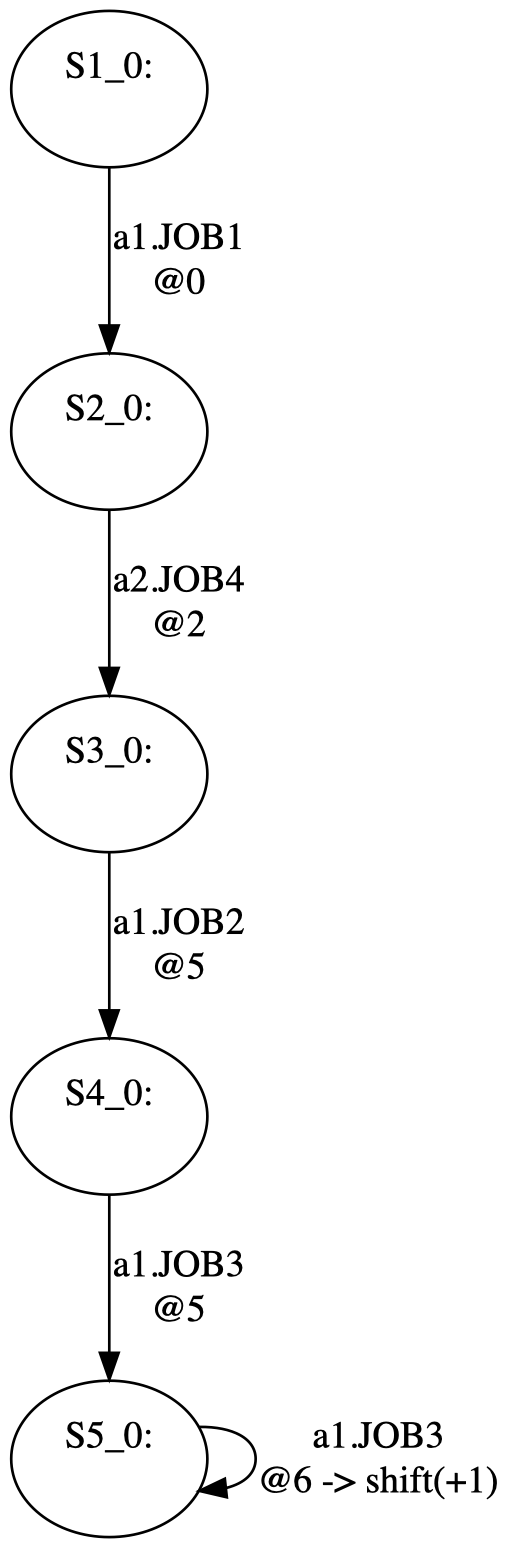}
  }
}

\caption{The state space of the Rebeca model of Listing~\ref{listing::rebeca-model}, generated by Afra using TTS and FTTS semantics}
\label{fig::two-states}
\end{figure*}




The verification algorithms of TTS with dense time are generally  PSPACE-complete as stated in \cite{DBLP:conf/rex/HenzingerMP91}. In the existing model checking tools commonly the properties are limited to a subset of TCTL properties without nested timed quantifiers. For this subset efficient algorthms are developed.
 In the case of Timed Rebeca, we use \textit{discrete time}, and hence TTS can be verified efficiently in polynomial time against TCTL properties. 
 Discrete time is the time model in which passage of time is modeled by natural numbers. We developed a model checking tool and a reduction technique for Timed Rebeca models based on TTS semantics against TCTL properties \cite{DBLP:journals/scp/KhamespanahKS18}. This toolset is not integrated in the Afra IDE \cite{afra-fsen}.

We also developed a tool for the model checking of Timed Rebeca models based on both TTS and FTTS semantics, which is integrated in Afra. The current implementation of the model checking toolset supports schedulability analysis, and checking for deadlock-freedom, queue-overflow freedom, and assertion based verification of Timed Rebeca models. 
Note that in FTTS, in each state actors may have different local clocks, so, writing meaningful assertion needs special care.
Assertions on state variables of one actor are not problematic.
The Timed Rebeca code of the case studies and the model checking toolset are accessible from Rebeca homepage \cite{Afra}.

Figure \ref{fig::two-states} shows the state space generated automatically by the model checking tool, Afra, for the Timed Rebeca model in Listing \ref{listing::rebeca-model} based on the two semantics, TTS and FTTS.
It is shown that the order of events are preserved while time progress and $\tau$ transitions are hidden. 
In state $S9\_0$ in Figure 2.a, and  state $S5\_0$ in Figure 2.b, you see how the transition system becomes bounded using a shift operation on the time.
The shift keyword means that  for example by the event \texttt{a1.JOB3}, we go back to state $S8\_0$ (or $S5\_0$), where all the values of state variables, local variables and messages in the  message buffers stay the same but the value of parameters related to time (including time tag of all messages and local clock value) change and have a shift by the same value.

\bibliographystyle{eptcs}
\bibliography{biblio}
\end{document}